\documentclass[preprint]{JASA}

\begin{document}

\title[]{Slowing and stopping the speed of sound}


\author{John L. Spiesberger}
\email{john.spiesberger@gmail.com}
\affiliation{Dept. of Earth and Environmental Science, U. of Pennsylvania, Philadelphia, PA 19104, USA}

\author{Ivesavega Djianto}
\email{ive@seas.upenn.edu}
\affiliation{U. of Pennsylvania, Philadelphia, PA 19104, USA}

\author{Justin Duong}
\email{juduong@sas.upenn.edu}
\affiliation{U. of Pennsylvania, Philadelphia, PA 19104, USA}

\author{Jisun Hwang}
\email{jisunh@sas.upenn.edu}
\affiliation{U. of Pennsylvania, Philadelphia, PA 19104, USA}

\author{Maria-Christina Nicolaides}
\email{marnic@seas.upenn.edu}
\affiliation{U. of Pennsylvania, Philadelphia, PA 19104, USA}

\author{Luke Stoner-Eby}
\email{lukeseby@seas.upenn.edu}
\affiliation{U. of Pennsylvania, Philadelphia, PA 19104, USA}

\author{Christian Stuit}
\email{cstuit@seas.upenn.edu}
\affiliation{U. of Pennsylvania, Philadelphia, PA 19104, USA}





\email{john.spiesberger@gmail.com}




\begin{abstract}
  Temporal interference between direct and surface-reflected paths induces large variation in the
  group speed of an acoustic signal between a source and receiver.    This speed goes to zero when the source and receiver
  approach one another and are within
  $c \tilde{\delta t}/2$ of the surface, where $c$ is the {\it in-situ}
  speed of sound and $\tilde{\delta t}$ is  the smallest temporal separation  between the paths at which interference
  initiates.
  At greater depths, the group speed can drop by many orders of magnitude. The effect diminishes far from a receiver
  as the size of the
  delay shrinks relative the overall time of propagation.
  The phenomenon is of great importance for methods designed to locate sounds via time differences of arrival (TDOA)
  as the group speed  between a sound and each receiver may  differ by orders of magnitude, a phenomenon that invalidates
  the geometrical interpretation of
  location by hyperboloids. Isodiachronic geometries are required to derive valid locations.
  Analogous to gravitational  black holes, where the speed of light is zero at the event horizon, ``three-dimensional acoustical black holes'' 
  can be present at acoustical receivers.
 
\end{abstract}


\maketitle



\section{\label{sec:1} Introduction}

Locations of underwater objects are often derived from measurements of emitted or reflected sounds.
Examples 
include ships, submarines, torpedoes, mines, fish, explosions, implosions, earthquakes, and calling animals.
Quite often, locations are derived from either the propagation time of sound or a TDOA between pairs of receivers. 
Propagation time and TDOA are converted to distance and  difference in distance from two receivers respectively by
multiplication by a speed of sound.
The origins of these methods date back at least to \citet{milne} where locations of underwater earthquakes were estimated from the
TDOA of tsunamis at multiple locations on land.  For three-dimensional spatial locations, the  ``three-dimensional effective speed'' is defined as,
\begin{equation}
c_{3d} \equiv l_1/t_m \ , \label{eq:c3d}
\end{equation}
where $l_1$ is the
distance of a straight-line  path from source to receiver and $t_m$ is the measured time of arrival. This is 
a required input for numerous software packages deriving location of calling marine mammals with  TDOA
\citep{mellinger_ishmael,mellinger_2024,gillespie_2008a,ces_code,conf_1,comp_localiz,baumgartner_2008}. 
Yet, something surprising occurred when  a  program was run to compute $c_{3d}$ to 
locate explosive sounds in the ocean.   
Values of $c_{3d}$ were 50 m/s less than the {\it in-situ} speed a few hundreds of meters  from the receiver.
Initially thinking the phenomenon was a software bug, it was later discovered the phenomenon was
caused
by interference between the direct and surface-reflected paths. The direct path is the shortest and does not reflect from a boundary.
The reflected path leaves the source, reflects once at the surface, then travels down to the receiver.
For sources near the receiver, the peak of the first energetic
arrival was delayed by interference.  This increased $t_m$ and caused $c_{3d}$ to decrease (Eq. \ref{eq:c3d}).
We have not found prior discussions of a large
drop of group speed due to this type of interference, though such a study may exist.
If it does, the phenomenon
is apparently not considered by contemporary acoustical oceanographers so is worth highlighting as its effects on the accuracy of location
can be important.  For example,  
a thorough investigation of this phenomenon demonstrates the $c_{3d}$ can drop to zero meters per second.
Using correct values for the $c_{3d}$ 
may be necessary to obtain a reliable confidence interval of location (CIL) for the class of models whose  inputs
require $c_{3d}$.

Interference effects are discussed across many scientific disciplines. 
For a single acoustic frequency, interference  near the  ocean's surface occurs due to the physics described by \citet{lloyd_mirror}.
However, propagation time is inherently a many frequency phenomenon since it
cannot be measured with single-frequency emissions.
For signals with positive bandwidth, interference affects propagation times of acoustic signals.  For example,
rays are a consequence of the interference of acoustical modes, each ray with its own time of propagation (e.g. Sec. 6.7.1, \citet{brek_lysanov}).
Interference is used to explain why the speed of light in  a  vacuum is slowed
by glass and water \citep{feynman_index_refraction} (Vol I,  Chapt 31).  When a gas is cooled near absolute zero, its interference
properties 
can be manipulated
to yield a  group speed of 17 m/s for light 
\citep{c_light_17}. Its speed can be decreased to zero meters per second by storing it in a quantum state
for  1 ms \citep{c_light_stop}.

The paper is organized as follows.
Sec. \ref{location_tdoa} discusses geometrical shapes useful for understanding how interference affects propagation speed and
how locations are understood when derived with TDOA.
Sec. \ref{three_d_speed} quantifies how the speed of signal  propagation is affected by interference between the direct and
reflected paths. Sec. \ref{examples} presents examples of the depressed speed along with quantification of its effects on the accuracy of locating sounds.
A discussion appears in Sec. \ref{discussion}.

\section{\label{location_tdoa} Locating objects with TDOA}

Deriving locations using TDOA can mainly be interpreted using one of two geometries: hyperboloids or isodiachrons \citep{isodiachrons}.
Both shapes are useful for understanding how interference affects location predictions as well as how it modifies the group speed of acoustic waves.

\subsection{\label{hyperbolas} Location via hyperbolas}

For two and three-dimensional models of location (2D and 3D), it is necessary to have at least four and five receivers respectively
to yield a mathematically unambiguous solution for
location from TDOA when measurements are made without error \citep{schmidt}.  The 
TDOA,
\begin{equation}
\delta t = t({\bf r}_2) - t({\bf r}_1) \ , \label{eq:delta_tm}
\end{equation}
is transformed into the source's difference in distance, $\delta d$, from receivers located at ${\bf r}_1 $ and ${\bf r}_2$  using
$ \delta d = c \delta t$, where $c$ is a constant wave speed. The locus of points in space sharing this difference 
defines a hyperbola in 2D space or hyperboloid in 3D space \citep{merriam_webster}. Adding receivers creates additional hyperbolas
or hyperboloids, whose intersections coincide at a single point with four receivers in 2D and and five receivers in 3D.
The constant speed  approximation can lead to large errors in practice so must be abandoned
to obtain a reliable CIL \citep{2d_black_holes,cse_eval}.

\subsection{\label{isodiachrons} Location via isodiachrons}

Deriving location from TDOA via hyperbolas requires multiplication of the TDOA by a constant speed (Sec. \ref{hyperbolas}).
  The isodiachron \citep{isodiachrons},  invented to account
for problems with a constant speed, is the locus of points for which the {\it difference in
  propagation time} is constant, allowing sound speed to differ between paths. Mathematically, its shape
is defined by,
\begin{equation}
\delta t = \frac{l_2}{c_2} - \frac{l_1}{c_1} \ , \label{eq:isodiachron_defn} 
\end{equation}
where $c_i$ is the group speed of propagation along path $l_i$ from source to receiver number $i$ and $l_i$ are usually taken to be
line segments.
When $c_1=c_2=c$,  Eq. (\ref{eq:isodiachron_defn})
reduces to $l_2-l_1= c \delta t$, and the isodiachron becomes a hyperbola. 
When $c_1 \neq c_2$, isodiachrons can look very different for realistic situations \citep{isodiachrons,2d_black_holes}.   When $c_i$ is
independent
of location, the isodiachron is class one. Otherwise it is class two \citep{isodiachrons}.

\section{\label{three_d_speed} 3D effective speed and interference}

\begin{figure}[ht]
  \centerline{\includegraphics[width=6in]{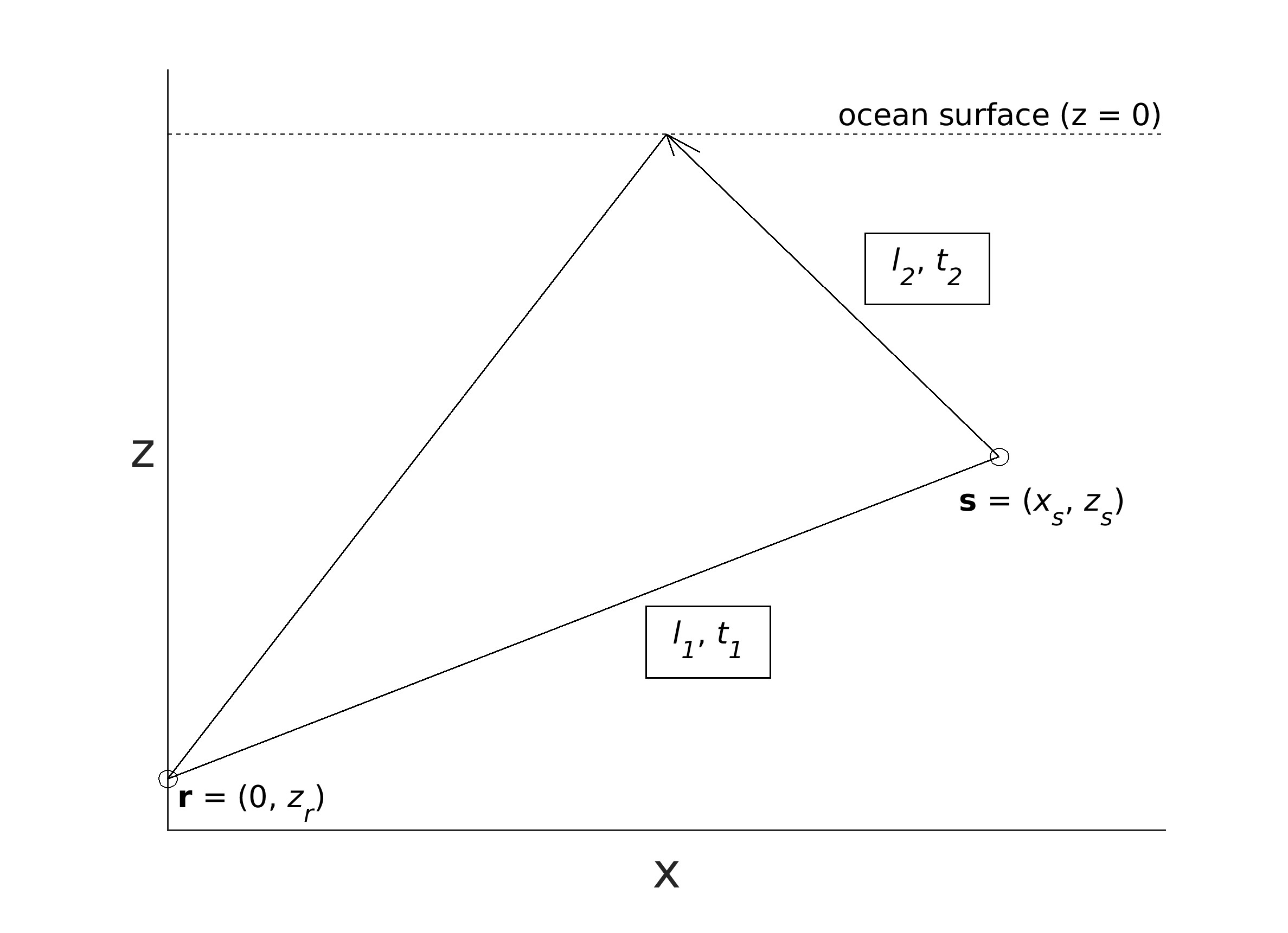}}
\caption{\label{fig:definition_fig} Sound propagates from source ${\bf s}$ to receiver ${\bf r}$ along direct and surface reflected paths
with lengths,  $l_1$ and $l_2$,  and propagation times, $t_1$  and $t_2$, respectively.}.
\end{figure}

The $c_{3d}$ is modified by interference over regions near an acoustic receiver, diminishing at large distances.
Analytical solutions for the effects of interference are derived below
for simple situations to gain intuition about where $c_{3d}$ is most impacted.

Because of Fermat's least-time principle (p. 457-463 of \citet{fermat}), the propagation time of a signal is affected by changes in path at second order,
and these are ignored with respect to a curved path.  Thus for short distances, the direct path is
assumed to be a line segment between source and receiver while the reflected path is assumed to be two line segments: one
from  source to the ocean's  surface, and  the other from the surface to the receiver.  The angle of ray incidence at the
surface is assumed to equal its angle of reflection. 

The Pythagorean distance between a receiver at ${\bf r} = (0,z_r)$ and source at ${\bf s} = (x_s,z_z)$ is,
\begin{equation}
l_1 = |{\bf s} - {\bf r}| =  \sqrt{x_s^2 + (z_r-z_s)^2} \ , \label{eq:l1}
\end{equation}
where the horizontal Cartesian axis is $x$ and vertical axis, $z$,  with
$z$ positive up and zero at the ocean's surface.    Similarly, the length of the surface reflected path is,
\begin{equation}
l_2 = \sqrt{x_s^2 + (z_r+z_s)^2} \ . \label{eq:l2}
\end{equation}
The z-weighted averages of sound speed along $l_1$ and $l_2$ are
$c_1$ and $c_2$ respectively. The propagation times along these paths are,
\begin{equation}
  t_i = \frac{l_i}{c_i} \ ; \ i=1,2 \ .  \label{eq:t}
\end{equation}
The difference in time between the measured and direct paths is,
\begin{equation}
\delta t_m = t_m-t_1 \ , \label{eq:dt}
\end{equation}
where $t_m$ is not usually equal to $t_2$ in the presence of interference.

\begin{figure}[ht]
  \centerline{\includegraphics[width=6in]{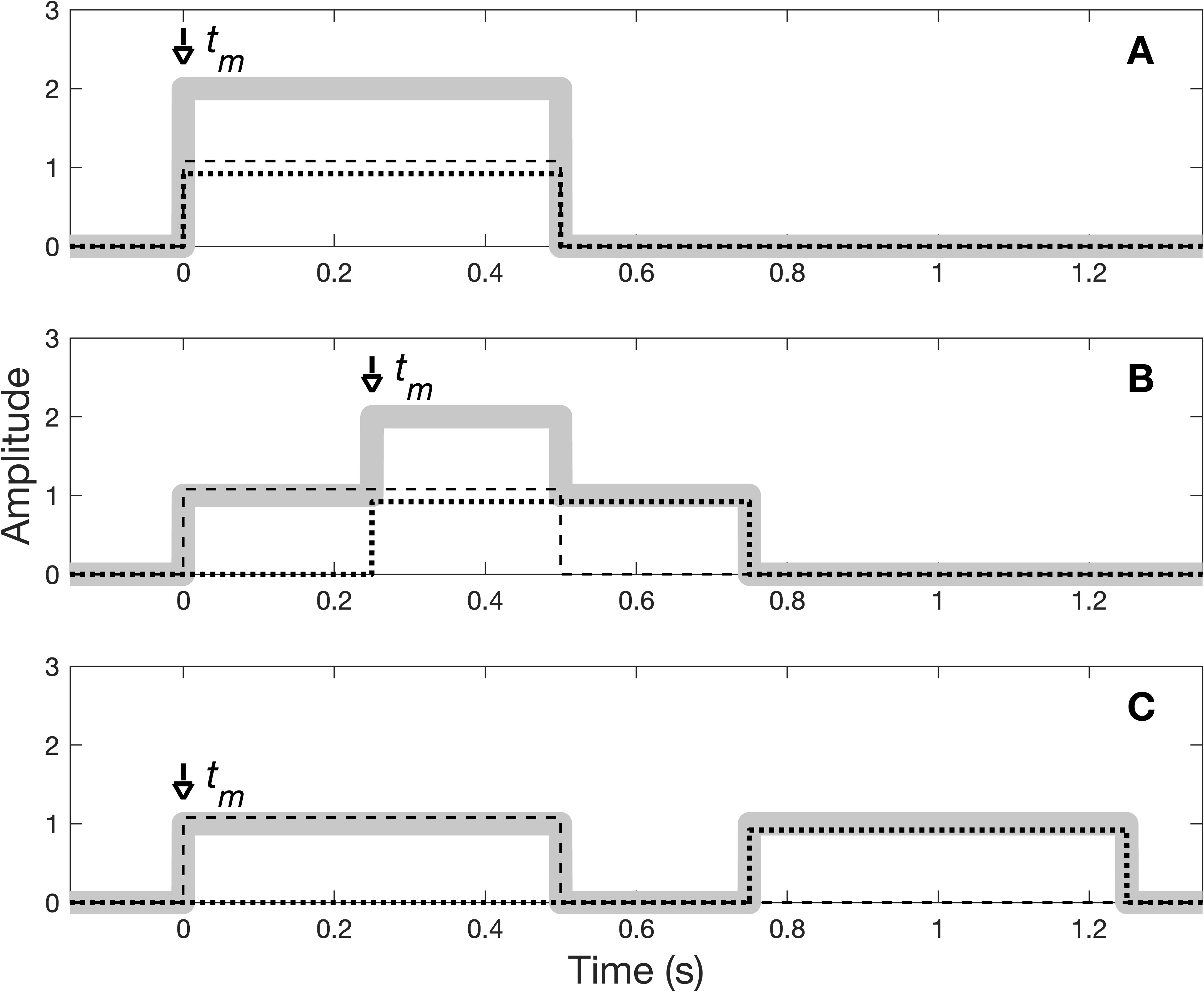}}
\caption{\label{fig:boxcar_fig}Measured times of arrival ($t_m$) based on relative arrival times of direct and reflected paths.
  Emitted waveforms are boxcars. {\bf A}: Direct (dashed) and reflected (dotted) arrive at same time (solid gray). {\bf B}: Same as {\bf A}
  but reflected arrives later. {\bf C}: Same as {\bf A} but reflected arrives much later so there is no interference.}.
\end{figure}

\begin{figure}[ht]
  \centerline{\includegraphics[width=3in]{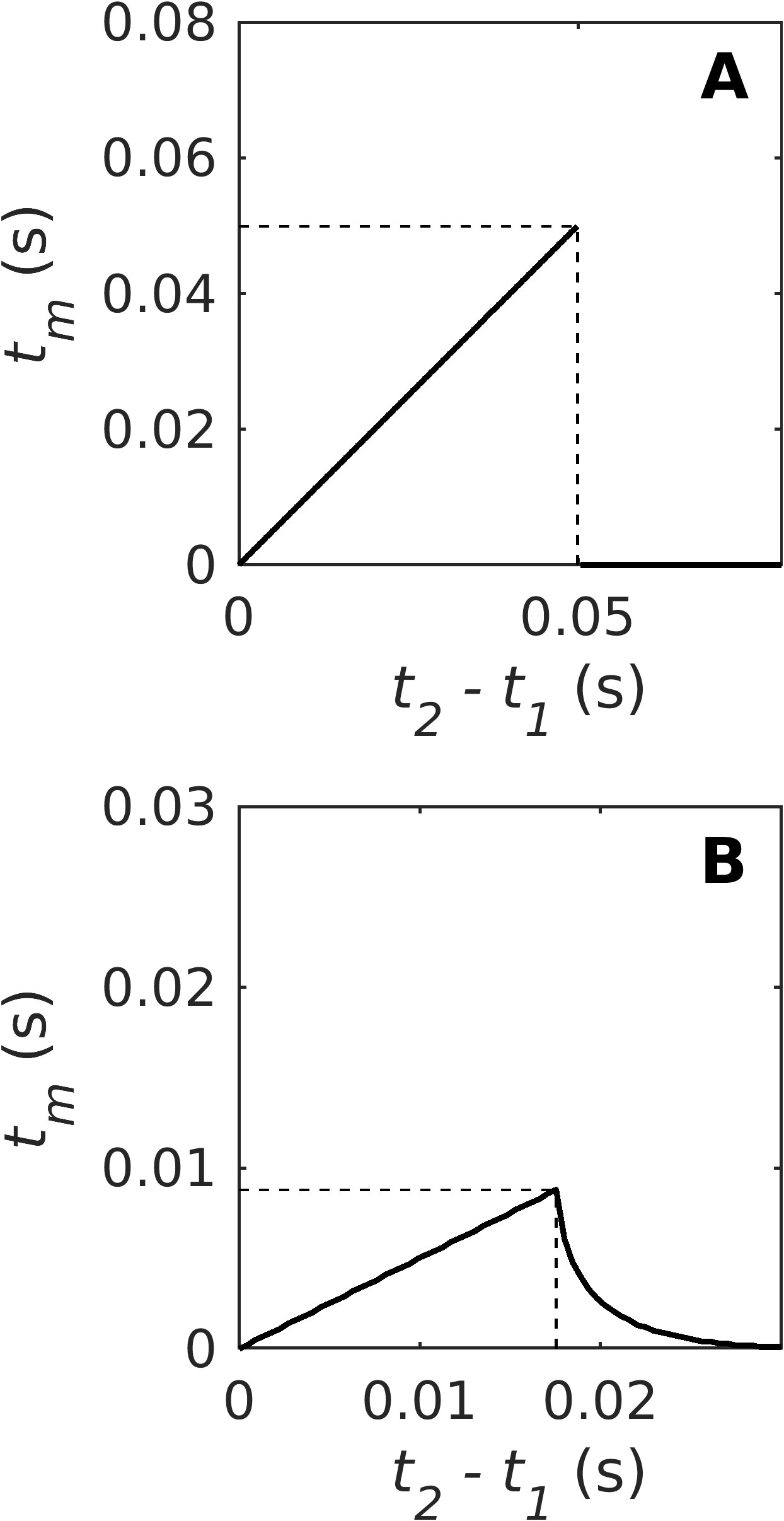}}
\caption{\label{fig:tm_vs_tdiff_figure} Pulse interference curves for boxcar ({\bf A}) and exponential pulses ({\bf B}). Pulse duration
  for boxcar is 0.05 s and $\tau=0.05$ s for exponential pulse in Eq. (\ref{eq:exp_pulse}).  Dashed
  lines indicate maximum value of $t_m$.}.
\end{figure}

Contours of constant $c_{3d}$  can be computed from Eq. (\ref{eq:c3d})
given specified waveforms and knowing $\delta t_m$'s  dependence
on $c_i$, ${\bf r}$, and ${\bf s}$. The surfaces on which $c_{3d}$ are constant have an  analytical
solution assuming $c_1$ and $c_2$  are the same, and the waveforms are boxcar functions
of duration $w$. The simplicity of the boxcar case stems from the reasonable assumption that
$t_m$ is taken to be the earliest time of  the
largest value of the sum of the two boxcar functions, when temporal interference occurs (Fig. \ref{fig:boxcar_fig}).  Thus,
$t_m=t_2$ (Fig. \ref{fig:tm_vs_tdiff_figure}A) and then,
\begin{equation}
\delta t_m = t_2-t_1 \ ; \ t_2-t_1 \leq w \ . \label{eq:dt_boxcar}
\end{equation}
When there is no temporal overlap,  $\delta t_m=0$ (Fig.  \ref{fig:boxcar_fig}). The function, $t_m(t_2-t_1)$ is called the
``pulse interference curve'' (Fig. \ref{fig:tm_vs_tdiff_figure}).

When interference occurs, Eq. (\ref{eq:dt_boxcar}) is substituted into Eq.
(\ref{eq:c3d}) along with values for $t_i$ from Eq. (\ref{eq:t}) to get,
\begin{equation}
c_{3d} = c \frac{l_1}{l_2} \ ; \ t_2-t_1 \leq w , \label{eq:c3d_boxcar_c_const}
\end{equation}
and where $c=c_1=c_2$ because the speed is spatially homogeneous. The surface on which
$c_{3d}$ is constant obeys,
\begin{equation}
\frac{l_1}{l_2} = \frac{c_{3d}}{c} = b = \mbox{constant} \ . \label{eq:c3d_surf1}
\end{equation}
Substituting Eqs. (\ref{eq:l1},\ref{eq:l2}) into Eq. (\ref{eq:c3d_surf1}) yields,
\begin{eqnarray}
  \frac{\sqrt{x_s^2 + (z_r-z_s)^2}}{\sqrt{x_s^2 + (z_r+z_s)^2}} &=& b \label{eq:b} \\ 
  (1-b^2)x_s^2 + (1-b^2)z_s^2 -2(1+b^2)z_r z_s &=& (b^2-1)z_r^2 \nonumber \\ 
  x_s^2 + z_s^2 - \frac{2(1+b^2)z_r}{(1-b^2)} z_s &=& -z_r^2  \nonumber \\ 
  x_s^2 + z_s^2 + g z_s&=&-z_r^2    \nonumber \\ 
 x_s^2 + (z_s+g/2)^2  &=& g^2/4 - z_r^2  \ ; \ b \neq 1 \  \label{eq:xs_boxcar_c_const} 
\end{eqnarray}
where,
\begin{equation}
g \equiv -\frac{2(1+b^2)z_r}{1-b^2} \ . \label{eq:e}
\end{equation}
This describes a circle with center $(0,-g/2)$ and radius,
\begin{equation}
  \rho=(g^2/4-z_r^2)^{1/2} = \frac{2 c c_{3d}}{c^2-c_{3d}^2} |z_r| . \label{eq:radius}
\end{equation}
Real-valued solutions exist when $\rho \ge 0$.  When $b=1$, there is no reflected path since 
Eq. (\ref{eq:c3d_surf1}) yields $l_1=l_2$.
Note $g \ge 0$ because  $b<1$ and $z_r \le 0$.

Eq. (\ref{eq:c3d_boxcar_c_const}) implies $c_{3d}$ approaches zero  when $l_1$ approaches zero,  subject to interference occurring. Since $c_{3d}$ can only
be zero when $l_1$ is zero, so the source and receiver must be co-located.  If they are too  deep, interference no longer occurs between the
direct and reflected  paths. The maximum  receiver depth
at which interference can occur with $c_{3d}=0$  equals,
\begin{equation}
  d_c = cw/2 \ ; \ \mbox{for boxcar pulses} \ , \label{eq:critical_depth_boxcar}
\end{equation}
and is called the critical depth. If the emitted pulse is not a boxcar function, $w$ is  replaced by,
\begin{equation}
  d_c = c \tilde{\delta t}/2 \ , \label{eq:critical_depth}
\end{equation}
where $\tilde{\delta t}$ is the minimum time delay, exceeding zero,
yielding insignificant interference between direct and reflected signals.

\subsection{\label{min_c3d_gt_0_section} Receiver below critical depth: minimum $c_{3d}$}

\begin{figure}[ht]
\centerline{\includegraphics[width=6in]{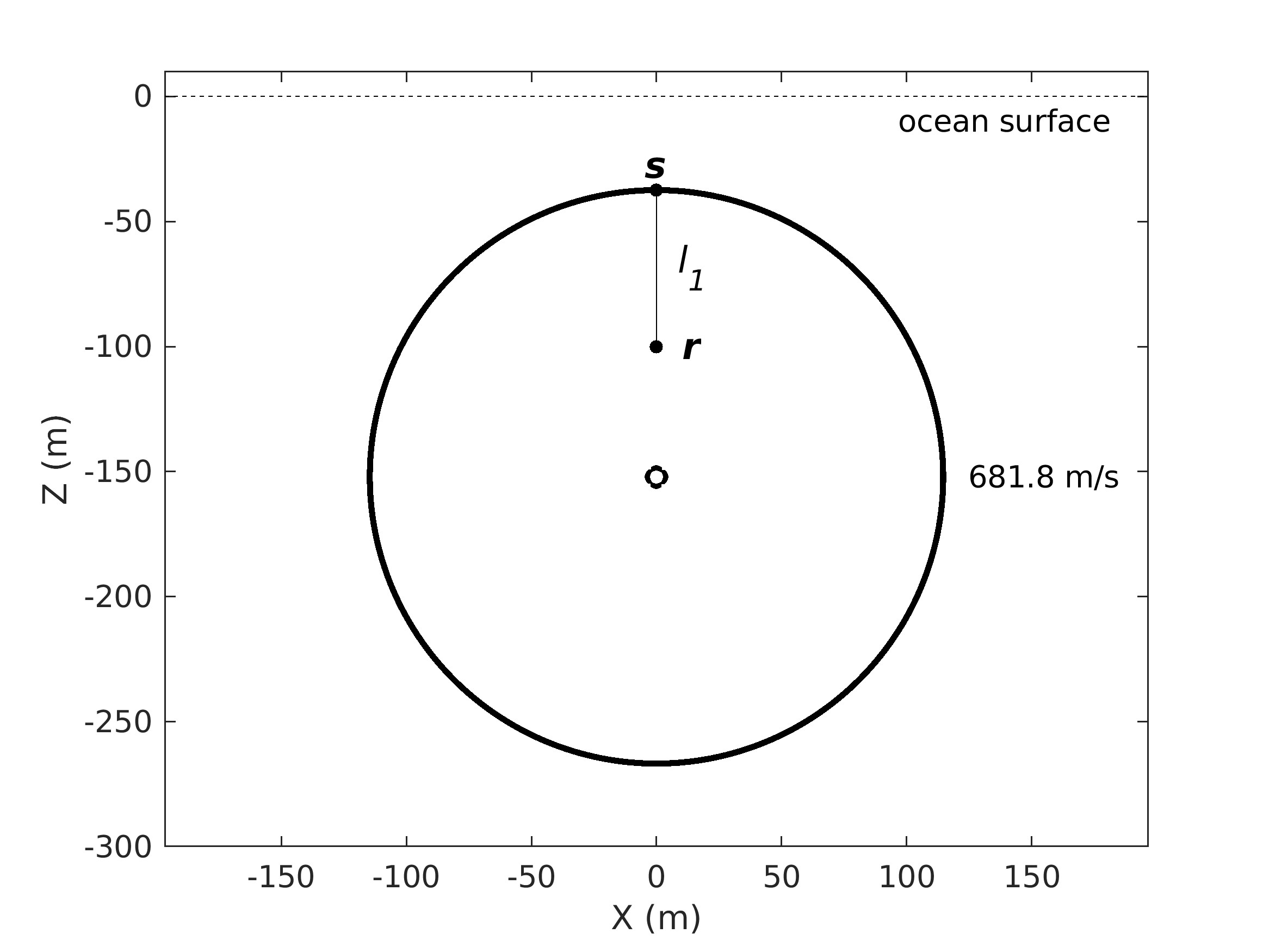}}
  \caption{\label{fig:c3d_min_const_c_analytical} Source and receiver at $(x,z)$ Cartesian coordinates (0,-37.5) and (0,-100) m respectively
  separated by distance $l_1$ in ocean with constant {\it in-situ} speed of sound of 1500 m/s. Source emits a boxcar function pulse
  of duration 0.05 s. Because receiver is below critical depth, $d_c=0.05 \ \mbox{s} \times 1500 /2 \ \mbox{m/s} = 37.5$ m,
  minimum possible effective speed, $\check{c}_{3d}$, exceeds zero and equals 681.8 m/s anywhere on circle (Eq. \ref{eq:c3d_min}).}
\end{figure}

When the {\it in-situ} speed of sound is a constant, $c$, the emitted pulse is a boxcar function, and the receiver's depth exceeds $d_c$,
$c_{3d}>0$ and its minimum, $\check{c}_{3d}$,  can be
derived analytically.  Surfaces
of constant $c_{3d}$ in the $x-z$ plane are circles (Eq. \ref{eq:xs_boxcar_c_const}).  The receiver is at $(0,z_r)$ and
the minimum speed occurs when $c_{3d} \equiv l_1/t_m$ is minimum, $\check{c}_{3d}$.  Thus, the location of the source is chosen
to minimize $l_1$ and maximize $t_m$.  The maximum delay between measured and direct path times is given by Eq. (\ref{eq:dt}) so
$t_m$ is maximum when $\delta t_m$ is $\widehat{\delta t}_m$, its maximum value.
So $\check{c}_{3d}$ corresponds to minimum $l_1$ and $\widehat{\delta t}_m$. occurring when,
\begin{equation}
l_2-l_1=c {\widehat \delta t}_m \ . \label{eq:l2_m_l1}
\end{equation}
The simplest geometry for solutions of $l_i$ is to place the source vertically above the receiver on the circle on which $\check{c}_{3d}$ occurs.
Then,
\begin{equation}
  l_1=z_s-z_r \ , \label{eq:l1a}
\end{equation}
because the source is above the receiver, and,
\begin{equation}
l_2=l_1 -2 z_s \ , \label{eq:l2a} 
\end{equation}
(Fig. \ref{fig:c3d_min_const_c_analytical}). Solving Eq. (\ref{eq:l2_m_l1}) for $l_2$ and substituting it into Eq. (\ref{eq:l2a}) yields
the vertical coordinate of the source,
\begin{equation}
  z_s = \frac{-c {\widehat \delta t}_m}{2} \ . \label{eq:zs_at_min}
\end{equation}
From Eq. (\ref{eq:l1a}), 
\begin{equation}
  l_1 = -c {\widehat \delta t}_m/2 - z_r \ . \label{eq:l1b}
\end{equation}
Since the measured time of arrival is,
\begin{equation}
  t_m = l_1/c + {\widehat \delta t}_m \ ,
\end{equation}
the minimum value of $c_{3d}$ is, 
\begin{equation}
  \check{c}_{3d} = l_1/t_m = -(c {\widehat \delta t}_m/2+z_r)/(l_1/c + {\widehat \delta t}_m) \ .
\end{equation}
Use $l_1$ from Eq. (\ref{eq:l1b}) in this to get,
\begin{equation}
  \check{c}_{3d} = c \frac{z_r + \widehat{\delta t}_m/2}{z_r - \widehat{\delta t}_m/2 } \ . \label{eq:c3d_min}
\end{equation}
When $c$ is not constant or when the emitted pulse is not a boxcar, $\check{c}_{3d}$ is computed numerically.

\subsection{\label{min_c3d_gt_0} Receiver below critical depth: interference boundary}

\begin{figure}[ht]
   \centerline{\includegraphics[width=3in]{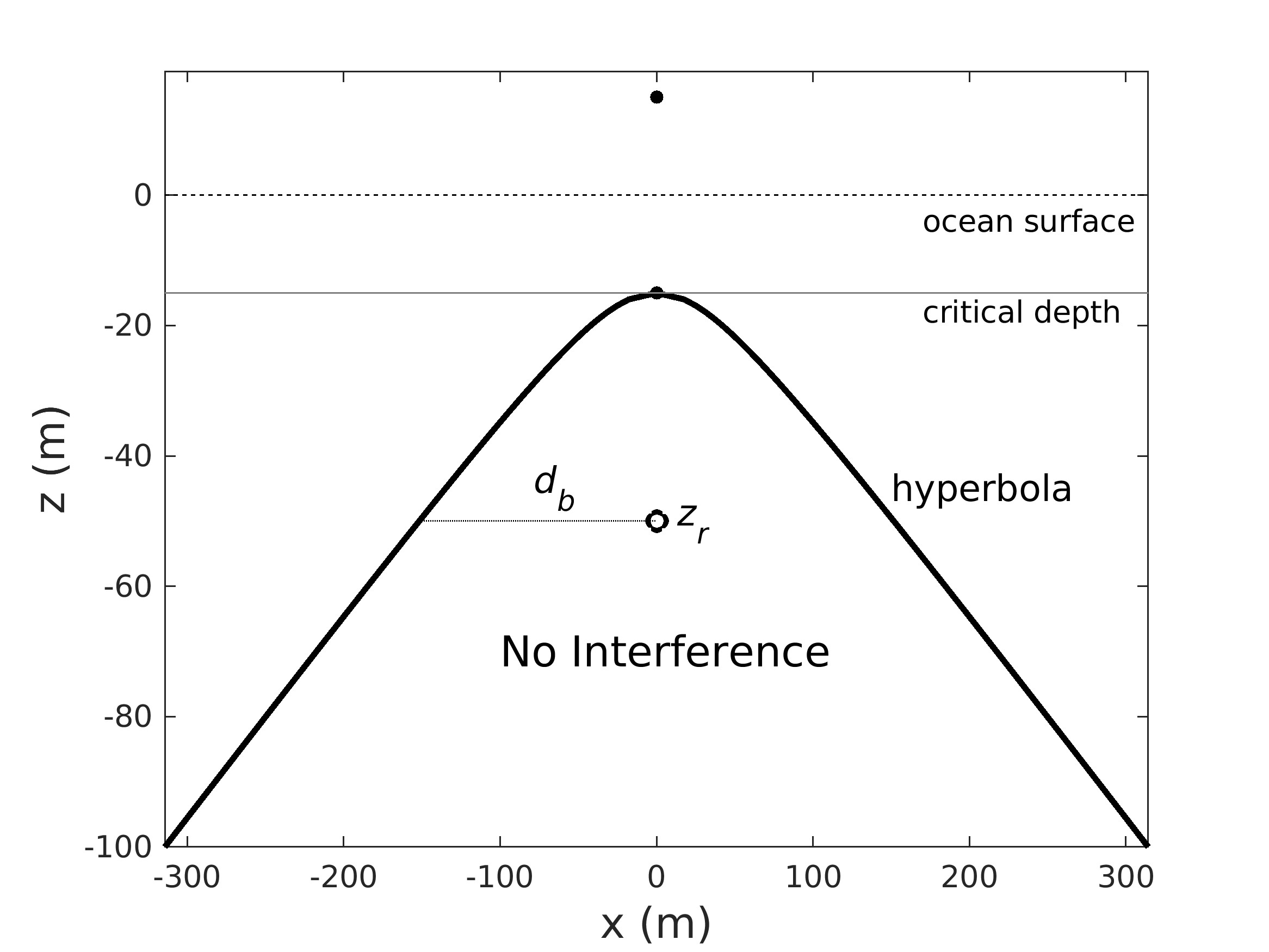}}
\caption{\label{fig:boundary_no_interference} Hyperbola separates regions where
    interference does and does not occur between direct and reflecting paths, assuming
    the {\it in-situ} speed of sound is 1500 m/s and the emitted pulse is a boxcar function of duration 0.02 s. Black dots are the hyperbola's foci.
    Receiver at 50 m depth is below the critical  depth, $d_c=wc/2 = 0.02 \ \mbox{s} \ \times \ 1500 \ \mbox{m/s} = 15$ m.
Horizontal distance to boundary, $d_b$, from Eq. (\ref{eq:d_b}).}.
\end{figure}

There are places where interference does not occur, and the ``interference-boundary'' separates them from
regions where interference occurs.
The boundary can be computed
analytically for a simple case.   It satisfies,
\begin{equation}
\frac{l_2}{c_2} - \frac{l_1}{c_1} = {\tilde \delta t} \ . \label{eq:interference_boundary} 
\end{equation}
This defines an isodiachron (Sec. \ref{location_tdoa}).
When $c_1=c_2=c$, Eq. (\ref{eq:interference_boundary})
reduces to $l_2-l_1= c {\tilde \delta t}$; a hyperbola (Sec. \ref{location_tdoa}, Fig. \ref{fig:boundary_no_interference}).
If the receiver is at or above  the critical depth, all regions are subject to interference.

When a receiver is below the critical depth, it is useful to compute the horizontal distance, $d_b$, between a vertical  line
passing through the receiver to the interference
boundary.   After many algebraic steps (not shown), Eq. (\ref{eq:interference_boundary}) yields,
%
%
\begin{equation}
 d_{b} = \sqrt{\frac{c^2 {\tilde \delta t}^2}{4} + \frac{4 z_r^2 z_s^2}{c^2 {\tilde \delta t}^2} -  z_s^2 - z_r^2  } \ \ ; \ z_r < -d_c \ \ z_s \leq d_c \ , \label{eq:d_b}
\end{equation}
where $d_c$ is defined in Eq.  (\ref{eq:critical_depth}) and $d_b$ is $x_s$ for this case. The horizontal distance, $d_b$, is zero
when $z_s = -c \tilde{\delta t}/2 = d_c$. For boxcar pulses, $c_{3d}$ is minimized at the interference boundary because $t_m$ is most delayed
when interference initiates (Fig. \ref{fig:boxcar_fig}). So, $d_b$ is the distance
to $\check{c}_{3d}$, where there is a discontinuity in the group speed. For exponential pulses, there is no discontinuity.
It was found empirically that replacing $\tilde \delta t$ with the relative time, $t_2-t_1$, corresponding to the maximum value
of $t_m$, in Eq. (\ref{eq:d_b}) yielded a distance 
that approximately located $\check{c}_{3d}$, and was well beyond the interference boundary (not shown).  

\subsection{\label{influence_distance} Interference and distance of influence}

Effects of interference on the $c_{3d}$ diminish at large distances from the receiver because the
corresponding time shift becomes small compared with the propagation time.
This is quantified using,
\begin{equation}
  \delta c \equiv c_1 - c_{3d}  = \frac{l_1}{t_1} -\frac{l_1}{t_m} \ , \label{eq:c3d_anomaly}
\end{equation}
where Eqs (\ref{eq:t}) and (\ref{eq:c3d}) are used for $c_1$ and $c_{3d}$.
When interference occurs, an analytical solution for $\delta c$ is derived when the {\it in-situ} speed of sound, c,
is constant and the emitted pulse is a boxcar function, so $t_m = t_2$.
Then
\begin{equation}
\delta c = l_1 \biggl( \frac{1}{t_1} - \frac{1}{t_2} \biggr) = \frac{l_1 \delta t}{t_1^2 (1+\frac{\delta t}{t_1})} \ . \label{eq:c3d_anomaly1}
\end{equation}
For larger  times of propagation, $\delta t/t_1 <  1$, and the Taylor series expansion for Eq. (\ref{eq:c3d_anomaly1}) is,
\begin{equation}
\delta c =    c^2 \frac{\delta t}{l_1}\biggl( 1 - \frac{\delta t}{l_1} c + (\frac{\delta t}{l_1} c)^2 + \cdots \biggr) \ , \label{eq:c3d_anomalyb}
\end{equation}
where $l_1 = c t_1$ was used. Retaining only the leading terms in $c \delta t/l_1$ yields an approximate distance of influence,
\begin{equation}
l_1 \sim c^2 \frac{\delta t}{\delta c} \ . \label{eq:dist_influence_approx}
\end{equation}
For example, let $c=1500$ m/s and $\delta t=$ 0.02 s. For $l_1=450$ m, $\delta c=100$ m/s and when $l_1=1125$ m, $\delta c=40$ m/s.
This shows the effect dies off with distance.

\subsection{\label{increased_c3d} Increased $c_{3d}$}

The $c_{3d}$ exceeds the direct paths' average speed when,
\begin{equation}
\frac{c_{3d}}{c_1} = \frac{t_1}{t_m} \ > 1 . \label{eq:c3_c1_ratio}
\end{equation}
In water, this can occur when the direct path travels through slow sound speeds associated with cold water and the reflected
path travels through fast sound speeds, associated with warm near-surface water.
It can also occur if the reflected path is replaced by a path traveling
from the source down into the solid Earth with large sonic speeds, then emerging back into the water to the receiver.

\section{\label{examples} Examples}

Some examples below model a pulse with an exponential shape following,
\begin{equation}
h(t) \equiv \exp [-(t-t_a)^2/(\tau/4)^2] \ , \label{eq:exp_pulse}
\end{equation}
where its time of arrival is $t_a$ and $\tau$ is a temporal scale. Fig. \ref{fig:tm_vs_tdiff_figure}B displays its  pulse interference
curve when the received amplitudes from both paths are equal.

\subsection{\label{c_const_1} $c_{3d}$  above and below critical depth}

\begin{figure}[ht]
\centerline{\includegraphics[width=8in]{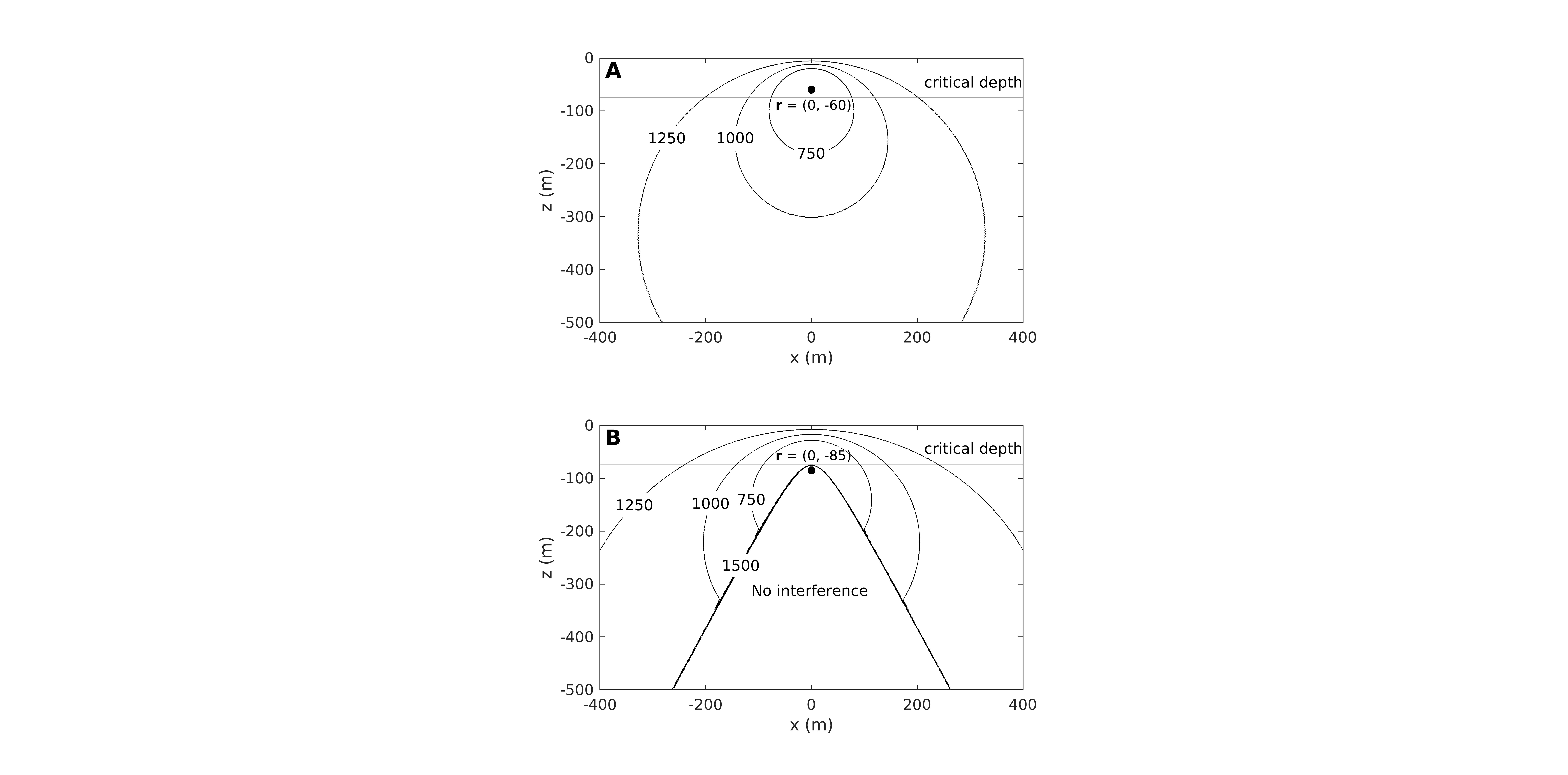}}
  \caption{\label{fig:c3d_contours} Contours of $c_{3d}$ (m/s) when {\it in-situ} speed of sound is 1500.1 m/s.
  {\bf A}: Receiver above critical depth.
  {\bf B}: Receiver below critical depth.  Hyperbola separates regions where interference does and does not occur.}
\end{figure}

\begin{figure}[ht]
  \centerline{\includegraphics[width=7in]{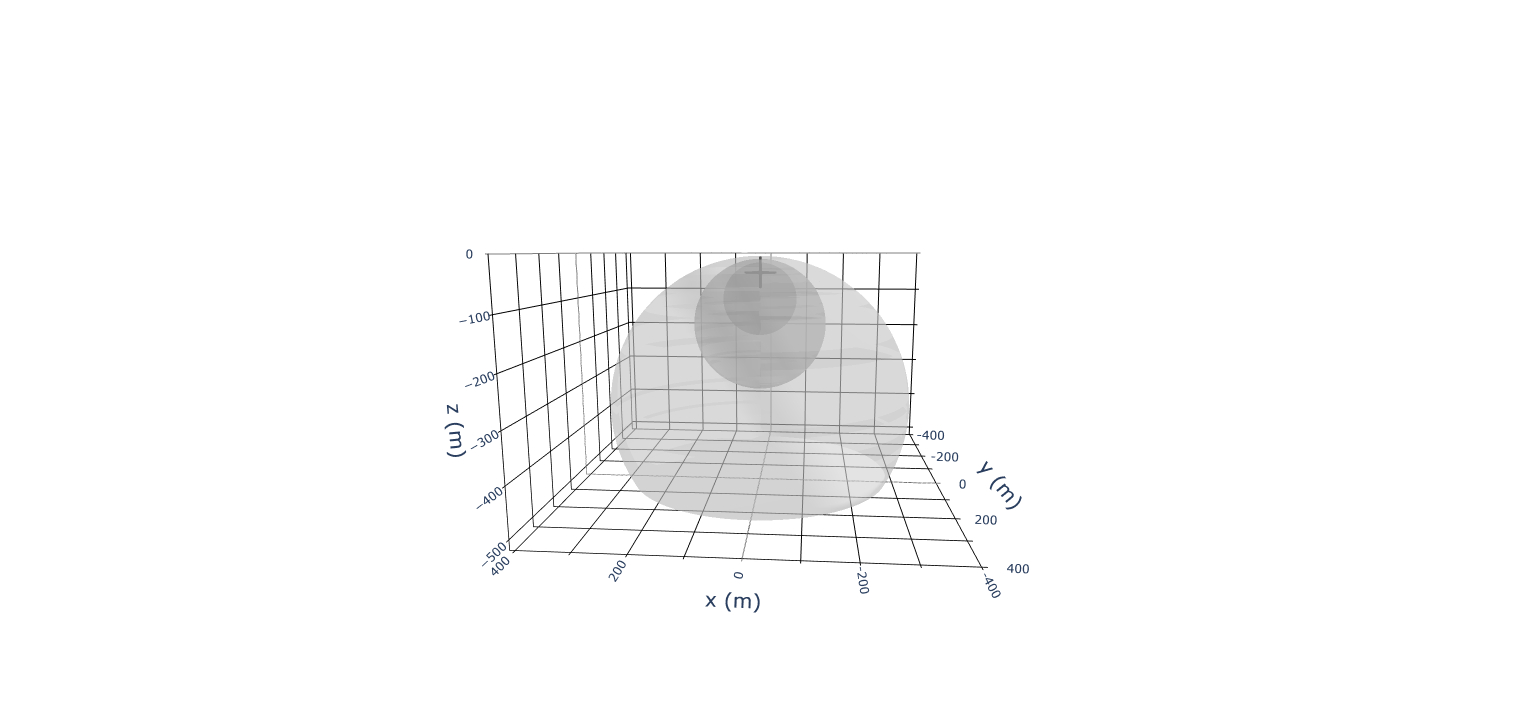}}
\caption{\label{fig:xyzc_level_sets_c_750_c_100_c_1250.png} Same as Fig. \ref{fig:c3d_contours} except rotated about vertical
  axis to level sets in three-dimensional space.  Receiver at cross and values of $c_{3d}$  are 750, 1000, and 1250 m/s from innermost to
outermost sphere.}
\end{figure}

The behavior of $c_{3d}$ is displayed when the {\it in-situ} speed of sound is a constant of 1500.1 m/s  and the emitted
pulse is a boxcar function of duration $w=0.1$ s.  The critical depth is 75 m (Eq.  \ref{eq:critical_depth}).
Consider a receiver depth of 60 m.
Since it is above the critical depth,
contours of constant $c_{3d}$ are circles (Eq. \ref{eq:xs_boxcar_c_const}) and approach zero as the distance
decreases between the source and receiver (Fig. \ref{fig:c3d_contours}A).  Rotating circular contours of constant $c_{3d}$ about the vertical axis
from Fig. \ref{fig:c3d_contours}A yields
spherical surfaces (Fig. \ref{fig:xyzc_level_sets_c_750_c_100_c_1250.png}).

Next consider a receiver at 75 m depth. Since it is below the critical depth,
a hyperbola separates the regions where interference does and does not occur (Sec. \ref{min_c3d_gt_0}).
Because the pulse shape is a boxcar function, there is a discontinuity in $c_{3d}$ at the interference boundary
where $c_{3d}$ goes from =1500.1 m/s to a small value depending on location on the interference boundary (Fig. \ref{fig:c3d_contours}B).

\begin{figure}[ht]
  \centerline{\includegraphics[width=5in]{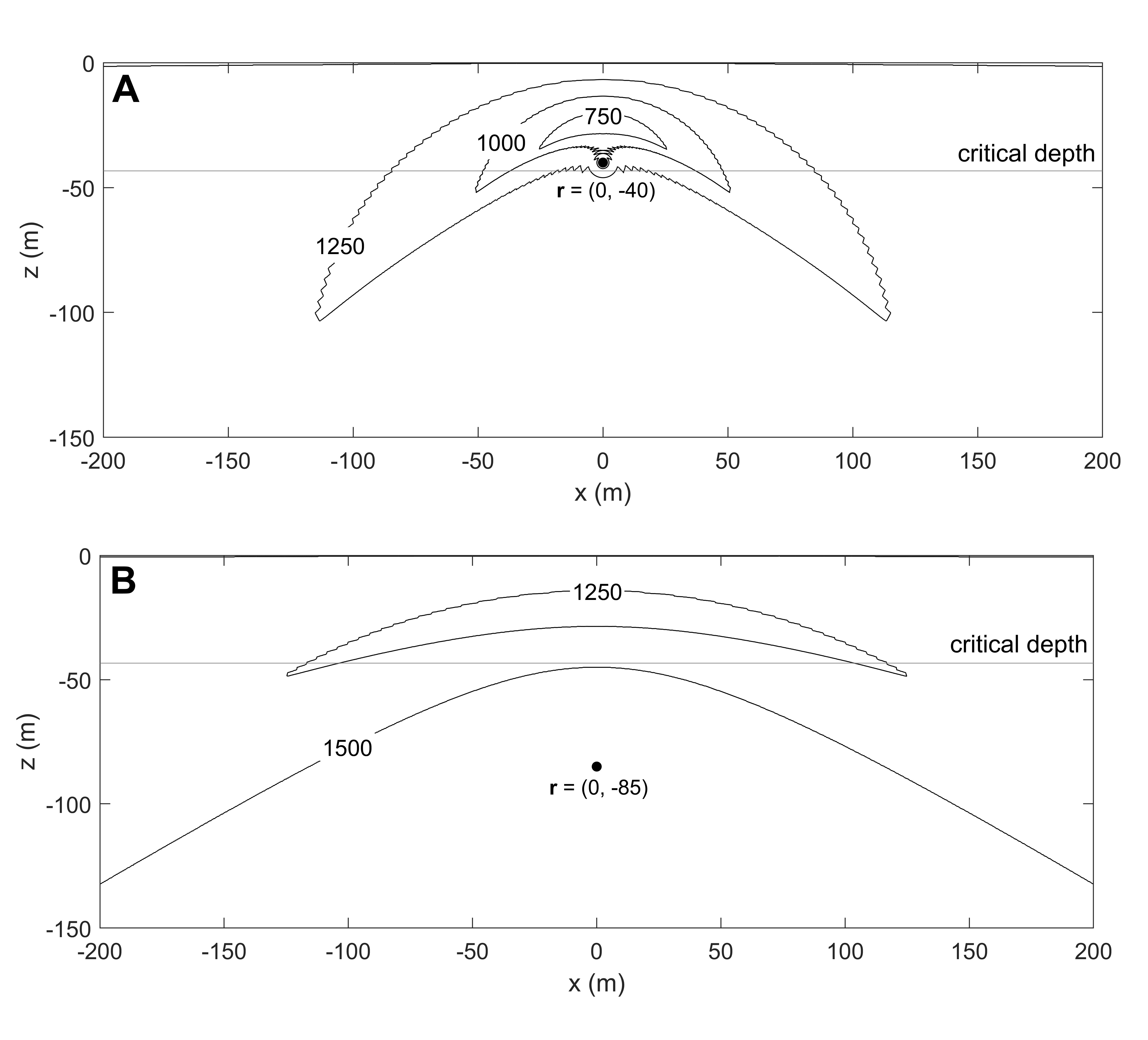}}
\caption{\label{fig:c3d_contours_exp} Contours of $c_{3d}$ (m/s) when {\it in-situ} speed of sound is 1500.1 m/s
  and transmitted pulse has exponential shape with  $\tau=0.1$ s in Eq. \ref{eq:exp_pulse}.
  {\bf A}: Receiver at $z=-40$ m is above critical depth.
  {\bf B}: Receiver at $z=-85$ m is below critical depth.}
\end{figure}

Unlike Fig. \ref{fig:c3d_contours} where a boxcar pulse is transmitted, the contours of constant
$c_{3d}$ are not circular when the transmitted pulse
is exponential.  For example, in  Fig. (\ref{fig:c3d_contours_exp}) the speed of sound is the same as Fig. \ref{fig:c3d_contours}
but the critical depth decreases from 75 m to 43.3  m. The minimum value of $z$ is changed from -500 m to -150 m
so details of the contours are visible.

\subsection{\label{constant_tdoa} Locations derived with TDOA}

\begin{figure}[ht]
  \centerline{\includegraphics[width=3in]{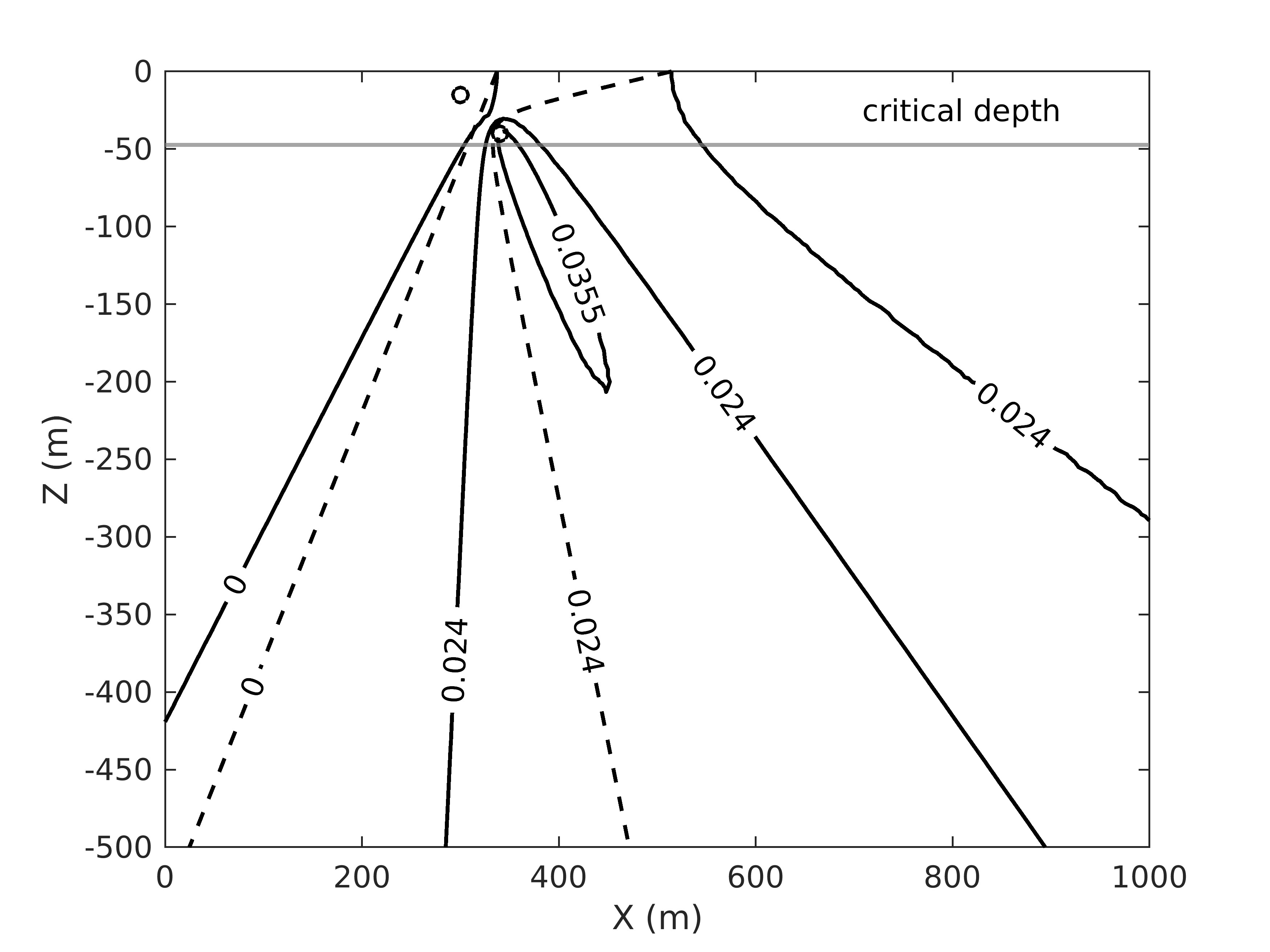}}
\caption{\label{fig:two_rec_xz_figure} Contours of measured TDOA ($\delta t_m$) (s) from  two receivers (circles)
  computed from dense grid of acoustic sources in
  Cartesian $(x,z)$ plane. Source emits exponential pulse (Eq. \ref{eq:exp_pulse}, $\tau=0.1$ s). $\delta t_m$ subject to
  interference between
  direct and reflected paths. {\it In-situ} speed of sound is 1500 m/s. Hyperbolas (dashed) compared with isodiachrons (solid lines) for same $\delta t_m$.
  Source cannot be located via hyperbolas when $|\delta t_m|$ exceeds
   propagation time of sound from first to second receiver but can be located with isodiachrons. Receivers above critical depth (gray line).}
\end{figure}

Locations of a source are estimated in a vertical plane with TDOA when the {\it in-situ} speed of sound is
$c=1500$ m/s and receivers are at Cartesian $(x,z)$ coordinates (300, -15) m and (340, -40) m (Fig. \ref{fig:two_rec_xz_figure}).
Measured TDOA are subject to interference between direct and reflected paths.  The emitted
pulse is exponential (Eq.  \ref{eq:exp_pulse}) with $\tau=0.1$ s. Both receivers are above the critical depth
equal to 47.32 m, determined numerically.
Measured TDOA are  contoured for values of 0 s, 0.024 s, and 0.0355 s.
Locations of hyperbolas are computed by multiplying the measured TDOA by 1500 m/s.  The largest possible TDOA for a hyperbola, 0.0314 s,
equals the distance between the receivers divided by the speed of sound. 
Next, these hyperbolas are compared with locations of the source derived with isodiachrons.

For the TDOA of 0 s, the hyperbola and isodiachron are similar,
differing at most by 50 m (Fig. \ref{fig:two_rec_xz_figure}).

For the TDOA of 0.024 s, the hyperbola and isodiachrons exhibit large differences. There are
{\it two} isodiachrons on which the sound could originate, and if the sound originated on the isodiachron
in the upper right side of Fig. (\ref{fig:two_rec_xz_figure}), the true location of the emitted sound  would
differ from the nearest corresponding hyperbola by about 500 m. This appears to be the first reporting that class two isodiachrons
can occur on separate surfaces.  The object could be anywhere on either of them. The multiplicity of isodiachrons is due to  the
fact that they are of class two. Class one isodiachrons do not exhibit this multiplicity \citep{isodiachrons}.

For the TDOA of 0.0355 s, there is no corresponding hyperbola because the largest possible TDOA
for a hyperbola is 0.0314 s.    Isodiachrons have no corresponding restriction,
and correct locations can be  derived.

\begin{figure}[ht]
  \centerline{\includegraphics[width=3in]{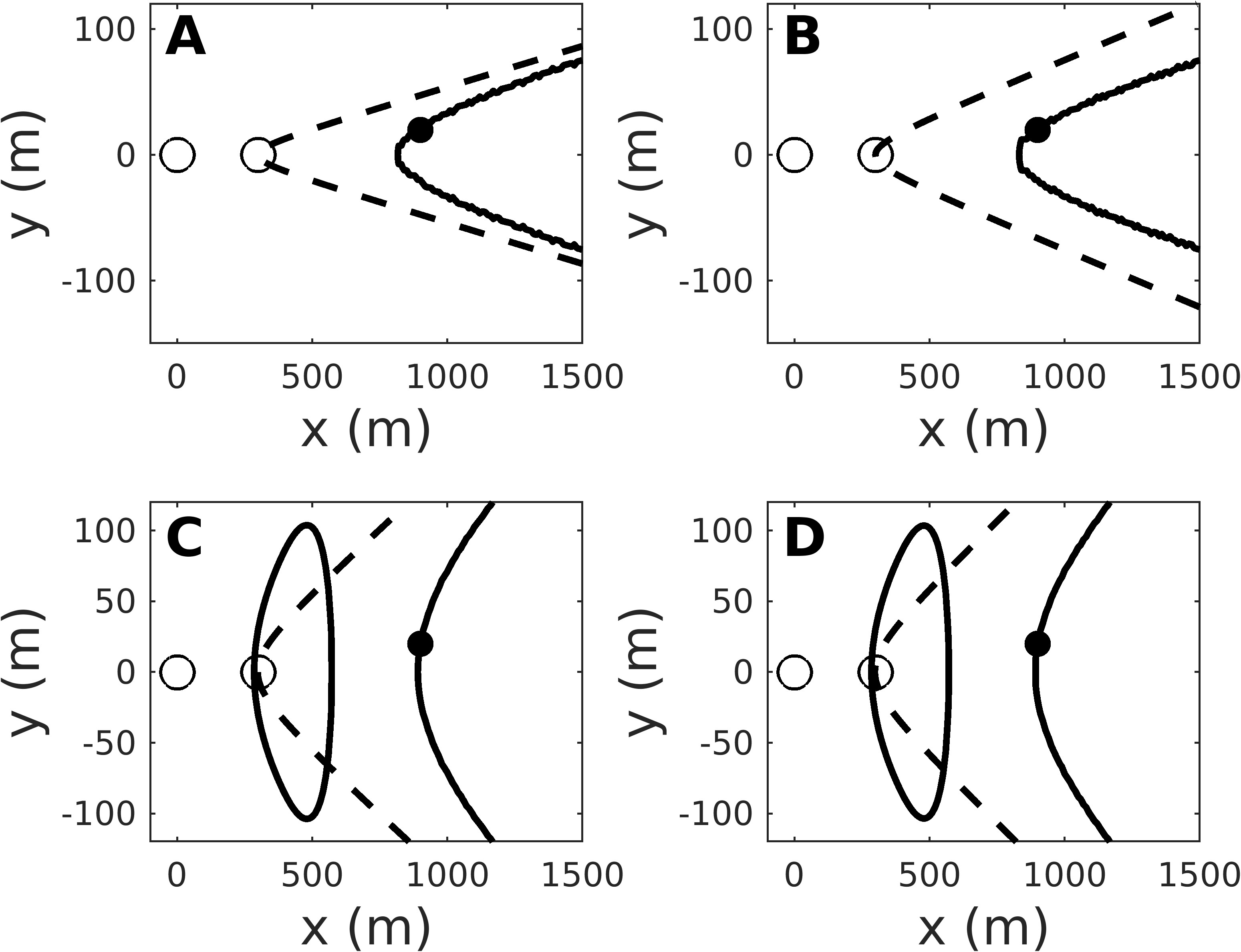}}
\caption{\label{fig:hyperbola_isodiachron}Hyperbola (dashed) and isodiachron (solid) for two receivers
  (open circles) and source (solid circle). {\bf A}: Receivers above critical  depth with {\it in-situ} sound speed of 1500 m/s. 
  {\bf B}: Same but sound speed derived from CTD data (Fig.  \ref{fig:ctd_plot}). Receivers above critical depth.
  {\bf C}: Same as {\bf A} but receivers at 90 m depth. These are below critical depth. {\bf D}: Same as {\bf C} but sound speed
  derived from CTD data (Fig.  \ref{fig:ctd_plot})}.
\end{figure}

\begin{figure}[ht]
  \centerline{\includegraphics[width=3in]{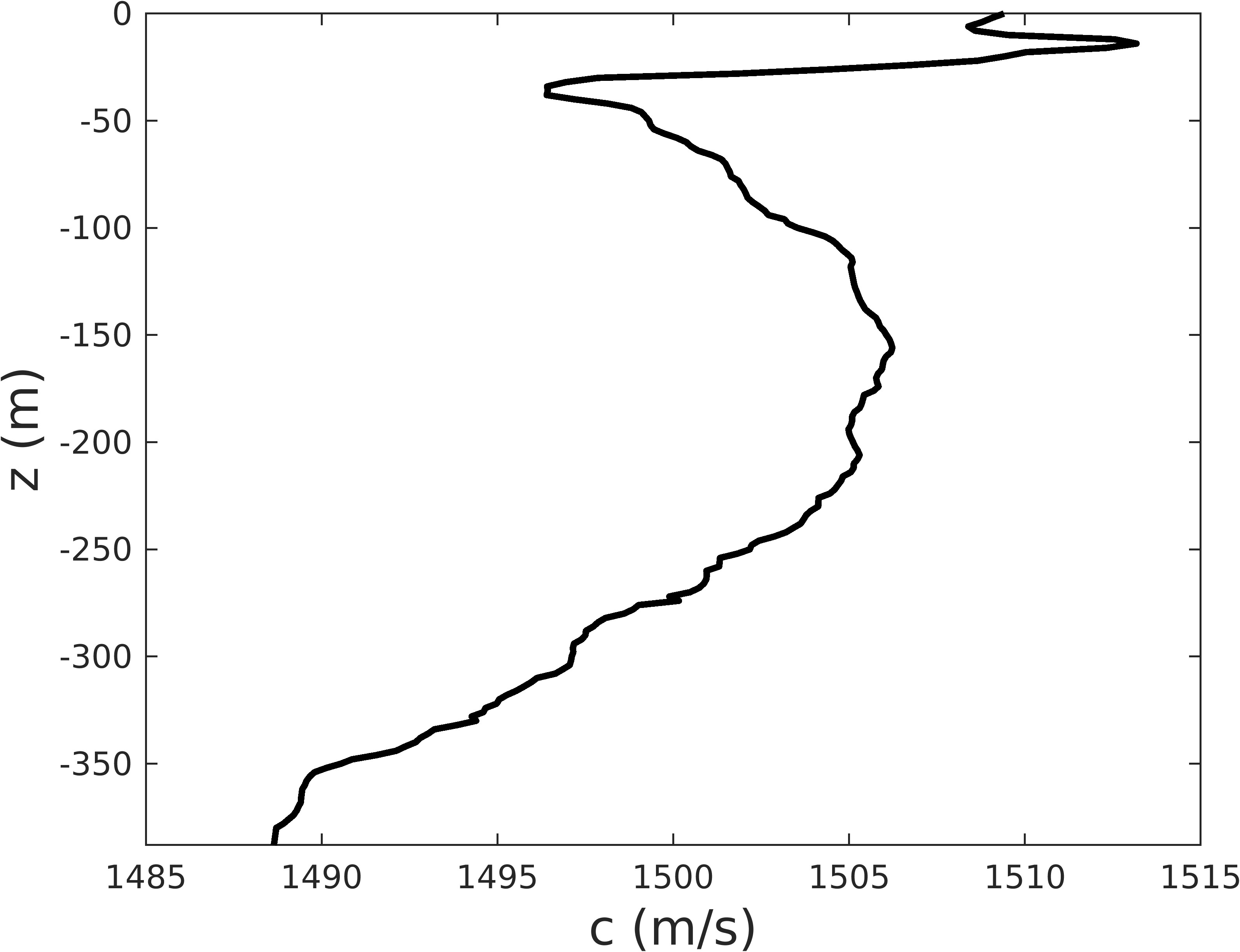}}
\caption{\label{fig:ctd_plot} Sound speed (m/s) at different depths. Derived from measurements of conductivity, temperature, and depth
  at 17:09 30 May, 2022 Greenwich Mean Time  
  at location 39.9608 $\mbox{}^{\circ}$N, 289.013 °E. Data courtesy Bill Hodgkiss, Scripps Institution of Oceanography.}
\end{figure}

Next, consider deriving the locations of a source in a horizontal  plane where the emitted sound is an exponential pulse
with $\tau=0.1$ s (Eq. \ref{eq:exp_pulse}). When the {\it in-situ} speed of sound is 1500 m/s, the  critical depth
is 47.5 m. Two receivers are placed at Cartesian $(x,y,z)$ coordinates (0,0,-30) and (0,300,-30) m and
source at (900,20,-30) m (Fig.  \ref{fig:hyperbola_isodiachron}).
The measured TDOA is multiplied by 1500 m/s to  yield
a hyperbola. It does not contain the source's location.   The source is 27.3 m  from the closest point on the hyperbola.
For the same measured TDOA, the isodiachron passes through the source. When the constant sound speed field is replaced
by measured sound speeds as a function of depth (Fig. \ref{fig:ctd_plot}), the measured TDOA is  multiplied by
1500 m/s to yield a hyperbola whose closest approach to the source is  46.1 m (Fig. \ref{fig:hyperbola_isodiachron}).
The isodiachron passes through the source because it utilizes different speeds along the paths to each receiver.
The receivers are above the critical  depth of  47.6 m.

The receivers and source are moved  from $z=-30$ to  $-90$ m, all below the critical depths for both the constant  and
measured sound speed fields. For the constant speed case, the hyperbola misses the source by 105.7 m  at closest approach
(Fig. \ref{fig:hyperbola_isodiachron}C).
There are two isodiachrons for this case, one of which passes through the source's location. The situation
is similar when the measured field of sound  speed is used  (Fig. \ref{fig:hyperbola_isodiachron}D). The source is missed by the hyperbola
by 112.9 m, but intersected by one of the isodiachrons.

\section{\label{discussion} Discussion}

Temporal interference between direct and reflecting paths can occur in solids, liquids, and gasses. This phenomenon potentially
leads to large changes in the group speed of acoustic signals. 
In water, this speed often drops by orders of magnitude within hundreds of meters of a receiver and even approaches zero
when the source and receiver are above the critical depth and near each other (Eq. \ref{eq:critical_depth}).
Many commonly-used methods for locating sounds require a single
speed of sound as an input to convert a TDOA to a  difference in distance from a pair of receivers, by which location
is geometrically interpreted as lying on a hyperbola or hyperboloid
\citep{mellinger_ishmael,mellinger_2024,gillespie_2008a,ces_code,conf_1,comp_localiz,baumgartner_2008}.  However, when a source is near one
receiver and  distant  from another, the group speeds to each receiver could be as different as 1 m/s and 1500 m/s.
This implies hyperbolas are inappropriate  for interpreting or deriving location.
Previous results based on these models should be reconsidered when the source might be near at least one receiver.

The isodiachron is an appropriate geometrical interpretation of location derived with TDOA when the speeds of propagation
differ between the source and each receiver \citep{isodiachrons}. This geometry was invented for the purpose
of deriving reliable locations in these circumstances.  Several  illustrations of their efficacy  here demonstrates
the true location of the  source lies on an isodiachron but not on a hyperbola where miss distances
range from 10 to 100 m (Figs. \ref{fig:two_rec_xz_figure},\ref{fig:hyperbola_isodiachron}).  Apparently, reliable CIL
need to account for the large decreases in group speed near the receivers.  At least one method using isodiachrons is designed to
generate reliable CIL in the presence of these variations of speed \citep{sbe,cse_eval}.

The intricacy of generating reliable CIL near receivers seems unintuitive because the action is all occurring
near the receivers where everything should seemingly be simple, but this is not so.
A similar complexity was recently discovered for 2D models of location where the vertical coordinate is discarded
and location is estimated with horizontal coordinates only.  The 2D effective speed of sound used to derive location
must vary with horizontal  distance from the receiver if a correct location is desired \citep{2d_black_holes}.
This speed goes to
zero meters per second when the source is above or below the receiver.  This
behavior is entirely caused by removing the vertical coordinate from the method of location.  The requirement
of zero speed above and below each receiver gave rise to the phrase, {\it 2D black holes}, to highlight
their importance in 2D models for location \citep{2d_black_holes}.
A narrated tutorial of 2D black holes and isodiachrons is available in
the supplementary material in \citep{cse_eval}.
In this paper, interference causes
the group speed to decrease to zero meters per second in three-dimensional  space near a receiver.
This is a real physical phenomenon and not a consequence of eliminating a vertical coordinate when deriving location.  In this sense,
there are acoustical black holes near receivers in 3D space when  interference  occurs between the direct and reflected paths.
This is an analogy to the naming of gravitational black holes where the speed of light  goes to zero
at the event horizon of a black hole.   We now see there are  both 2D and 3D acoustical black holes.

\begin{acknowledgments}
  Research  was supported by Office of Naval Research (ONR) grant  N00014-23-1-2336 and the University of Pennsylvania
  via the Penn Undergraduate Research Mentoring Program.
  The CTD data were provided through the ONR sponsored experiment SBCEX22. We thank Maya Mathur and Mary Putt for their comments.
\end{acknowledgments}

\section{\label{declarations} Author Declarations}
Sequential bound estimation and its related technologies is a commercial location service. No other
conflicts of interest.

The data that support the findings of this study are available within the article
and its supplementary material.








\bibliography{sampbib}


\end{document}